
\input phyzzx
\author{Shin'ichi Nojiri}
\address{Department of Mathematics and Physics}
\address{National Defense Academy}
\address{Hashirimizu, Yokosuka 239, JAPAN}
\author{Ichiro Oda}

\tolerance=10000

\def\cref#1{\rlap,\attach{#1}}
\def\pref#1{\rlap.\attach{#1}}
\def\refmark#1{\attach{#1}}\def\refmark#1{\attach{#1}}

\def\ie{{\it i.e.}}

\def\LA{\ \Longrightarrow\ }
\def\e{{\rm e}}

\def\ln{{\rm ln\,}}
\def\sinh{{\rm sinh\,}}
\def\tanh{{\rm tanh\,}}
\def\half{{1\over 2}}

\def\pint{{1 \over 2\pi}\int d^2x\,}
\def\sqg{\sqrt{-g}}
\def\ephi{\e^{-2\phi}}
\def\erho{\e^{-2\rho}}
\def\epphi{\e^{2\phi}}
\def\eprho{\e^{2\rho}}
\def\ephirho{\e^{2(\phi-\rho)}}
\def\RN{Reissner-Nordstr\"om}
\def\RST{Russo, Susskind and Thorlacius}
\def\txz{\tilde x_0}
\def\om{\ephi+{N \over 48}\phi}

{\hsize=17.5truecm \leftskip=9.5cm
{NDA-FP-5/92, OCHA--PP-27}\par
\vskip -4mm
{July 1992}\par}
\title{Semi-classical Approach to Charged Dilatonic Black Hole
in Two Dimensions}
\author{Shin'ichi Nojiri}
\address{Department of Mathematics and Physics}
\address{National Defense Academy}
\address{Hashirimizu, Yokosuka 239, JAPAN}
\author{Ichiro Oda}
\address{Faculty of Science, Department of Physics}
\address{Ochanomizu University}
\address{1-1 Otsuka 2, Bunkyo-ku, Tokyo 112, JAPAN}

\endpage

\abstract{We consider exactly solvable semi-classical theory of
two dimensional dilatonic gravity with electromagnetic interactions.
As was done in the paper by Russo, Susskind and Thorlacius,
the term which changes the kinetic term is added to the action.
The theory contains massless fermions as matter fields and
there appear the quantum corrections including chiral anomaly.
The screening effect due to the chiral anomaly has a tendency to
cloak the singularity.
In a region of the parameter space, the essential behavior of the
theory is similar to that of Callan, Giddings, Harvey and
Strominger's dilatonic black hole theory modified in the paper
by Russo, Susskind and Thorlacius and the singularity formed by
the collapsing matter emerges naked.
We find, however, another region of the parameter space where
the singularity disappears in a finite proper time.
Furthermore, in the region of the parameter space,
there appears a discontinuity in the metric on the trajectory
of the collapsing matter, which would be a signal of topology
change.}

\endpage
\REF\i{S.W. Hawking\journal Comm.Math. Phys. &43 (75) 199}
\REF\v{G. 't Hooft\journal Nucl.Phys. &B335 (90) 138}
\REF\ii{Y. Aharonov, A. Casher and S. Nissinov\journal Phys.Lett.
&B191 (87) 51}
\REF\iv{C.F.E. Holzhey and F. Wilczek, preprint IASSNS-HEP-91/71}
\REF\iii{C.G. Callan, S.B. Giddings, J.A. Harvey
and A. Strominger\journal Phys.Rev. &D45 (92) R1005}
\REF\vi{J.G. Russo, L. Susskind and L. Thorlacius,
preprint SU-ITP-92-4 (1992)}
\REF\vii{L. Susskind and L. Thorlacius,
preprint SU-ITP-92-12 (1992)}
\REF\xi{S.W. Hawking, preprint CALT-68-1774}
\REF\xii{T. Banks, A. Dabholkar, M.R. Douglas
and M. O'Loughlin\journal Phys.Rev. &D45 (92) 3607}
\REF\xviii{A. Bilal and C.G. Callan, preprint PUPT-1320 (1992)}
\REF\xix{S.P. de Alwis, preprints, COLO-HEP-280, 284 (1992)}
\REF\xx{K. Hamada, preprint UT-Komaba 92-7}
\REF\xvii{J.G. Russo, L. Susskind and L. Thorlacius,
preprint SU-ITP-92-17 (1992)}
\REF\xvi{S. Nojiri and I. Oda, preprint NDA-FP-5/92, OCHA-PP-26
(1992)}
\REF\viii{N. Ishibashi, M. Li
and R. Steif\journal Phys.Rev.Lett. &67 (91) 3336}
\REF\ix{M.D. McGuigan, C.R. Nappi
and S.A. Yost\journal Nucl.Phys. &B375 (92) 421}


\chapter{Introduction}
One of most important theme in elementary particle physics is to
construct the quantum theory of the gravity.
Black hole evaporation, however,  provides a serious problem.
A quantum mechanically
pure state which describes gravitationally collapsing matter to form
a black hole, evolves into a mixed quantum state which describes
Hawking radiation\pref\i
This implies that there does not exist a well-defined S-matrix theory.
There are several scenarios to solve this problem of the loss of
quantum informations.
Hawking has given a most radical proposal that the black hole completely
evaporates and the quantum coherence is lost in the gravitational
collapse.
On the other hand, 't Hooft has proposed that Hawking radiation
carries off informations about the quantum states of the black
holes\pref\v
Another conservative proposal is that the process of the collapse
and the radiation leaves a stable remnant which carries the
information of the initial configuration of the system\pref\ii
However, if the remnant has the mass of order the Planck scale,
gravitational effects would produce the light remnants in pair and
the lifetime of stars would be shorter than observed.
Therefore the mass of the remnants should be macroscopic.
A candidate of such remnants is the extremal black holes of
\RN\ solutions\pref\iv
The extremal black holes have vanishing temperatures and the space-time
metric is not singular everywhere.

Recently, Callan, Giddings, Harvey and Strominger (CGHS) have
investigated an interesting toy model of two dimensional gravity\pref\iii
In this model, the gravity couples with a dilaton and conformal
matter fields. They found classical exact solutions, including the
solutions describing the formation of a black hole by collapsing
matter.
Furthermore Hawking radiation and the back reaction of the metric
can be described by adding correction terms to the classical
equations of motion.
The correction terms derive from the conformal anomaly of the matter
fields.
It was also argued that the quantum correction would remove the usual
black hole singularity.
Several groups\cref{\vi - \xii} however, pointed that gravitational
collapse always leads to a curvature singularity.
Susskind and Thorlacius have found massive static solutions with
zero temperature but they have shown that a decaying black hole
cannot evolve into one of them\pref\vii

Although the original CGHS semi-classical equations of motion have
not been solved explicitly, Bilal and Callan\refmark\xviii and
de Alwis\refmark\xix have found that the original model can be
modified to allow obtaining explicit exact solutions of quantum
black holes.\foot{
Similar but different quantum model was proposed by Hamada\pref\xx}
After that, \RST\ have studied a variation\pref\xvii
They have changed a kinetic term by adding the following term to the
action,
$$\delta S=-{\kappa \over 4\pi}\int d^2x\sqg \phi R\ . \eqn\qi$$
Here $\phi$ is a dilaton field and $R$ is a scalar curvature.
Then there appears a conserved current $J^\mu$, which has the
following form,
$$J^\mu=\partial^\mu (\phi-\rho)\ , \eqn\qii$$
in the conformal gauge,
$$g_{\mp\pm}=-\half\eprho\ , \ \ g_{\pm\pm}=0\ . \ \eqn\eii$$
The current has already appeared in CGHS's classical theory.
The conservation law
$\partial_\mu J^\mu=\partial_\mu\partial^\mu(\phi-\rho)=0$
allows to fix the residual gauge symmetry of reparametrization by
choosing the condition $\phi-\rho=0$, which is given by a sum of
holomorphic function $w^+(x^+)$ and anti-holomorphic function
$w^-(x^-)$ in general : $\phi(x^+,x^-)-\rho(x^+,x^-)=w^+(x^+)+w^-(x^-)$.
The gauge choice is very useful when we construct explicit solutions
of classical and semi-classical equations of motion.

In the previous paper\cref\xvi the authors have coupled electromagnetic
fields to CGHS dilatonic gravity.
The coupling has been chosen so that the current \qii\ is conserved.
This model can be also solved classically.
Among the solutions describing static black holes,
there exist extremal solutions which have zero temperatures.
In the extremal solutions, the space-time metric is not singular
and these solutions should be analogues of four dimensional extremal
\RN\ black holes.
We also obtain the solutions describing charged matters (massless
fermions) collapsing into black holes.
Through the collapse, not only future horizon but past horizon
is also shifted.
The quantum corrections have been also discussed by including chiral
anomaly.
When the matter collapses into a black hole,
except the case that the resulting black hole is extremal,
the curvature singularity also appeared in the quantum theory
in a way similar to CGHS model.

In this paper, we consider exactly solvable semi-classical theory
of two dimensional dilatonic gravity with electromagnetic
interactions. The term corresponding to \qi\ is added to the action.
The screening effect due to the chiral anomaly has a
tendency to cloak the singularity.
In a region of the parameter space, the behavior of the
theory is similar to that of modified CGHS's dilatonic black hole
theory which has been analyzed by Russo, Susskind and
Thorlacius\refmark\xvii and the singularity formed by
the collapsing matter emerges naked.
In another region of the parameter space, however, the singularity
disappears in a finite proper time.
Furthermore, in the region of the parameter space,
there appears a discontinuity in the metric on the trajectory
of the collapsing matter, which would be a signal of topology
change.

In the next section, we review the classical theory of charged
dilatonic black holes based on the previous paper.
In section 3, we consider the semi-classical theory
by adding the correction terms to the classical action.
The correction terms contain the terms which come from the
conformal and chiral anomalies and the term corresponding to \qi .
The last section is devoted to summary and discussion.

\chapter{Classical Theory of Charged Dilatonic Black Holes}

We start with the following action,
$$S_c=\pint\sqg\{\ephi(R+4(\nabla \phi)^2+4\lambda^2)
-{\e^{a\phi} \over g_A^2}F^2
-\sum_{j=1}^N i\bar\Psi_j\gamma^\mu(D_\mu-iA_\mu)\Psi_j\}
\ . \eqn\ei$$
Here $\Psi_j={\psi^L_j \choose \psi^R_j}$'s are $N$ massless complex
fermions. $g_A$ is a $U(1)$ electromagnetic gauge
coupling constant and $D_\mu$ is a covariant derivative.
If we fix $a=-2$, this action, except the fermion part, describes
an effective field theory derived from string theory\pref{\viii, \ix}
In this paper, we choose $a=2$, which gives a conseved current \qii\
and we can obtain the classical exact solutions of this model.
When we fix the gauge freedom of reparametrization invariance
by the conformal gauge \eii\ and that of $U(1)$ gauge symmetry by
the following light-cone gauge condition
$$A_-=0 \ , \eqn\eiib$$
the action \ei\ reduces into
$$\eqalign{S_c&=\pint\{\ephi(4\partial_+\partial_-\rho
-8\partial_+\phi\partial_-\phi+2\lambda^2\eprho)\cr
&\ \ \ \ +{4 \over g_A^2}\e^{2\phi-2\rho}F_{+-}^2
+{i \over 2}\sum_{j=1}^N(\psi^{L*}_j\partial_-\psi^L_j
+\psi^{R*}_j(\partial_+-iA_+)\psi^R_j)\}
\ .} \eqn\eiii$$
Here $F_{+-}=-\partial_-A_+$.
The metric equations are given by
$$\eqalign{0&=T_{++}=\ephi
(4\partial_+\rho\partial_+\phi-2\partial_+^2\phi)
+{i \over 4}\sum_{j=1}^N
(\psi^{L*}_j\partial_+\psi^L_j-\partial_+\psi^{L*}_j\psi^L_j)\cr
&\ \ \ \ \ \ \ +\half A_+\sum_{j=1}^N\psi^{L*}_j\psi^L_j \ ,
\cr
0&=T_{--}=\ephi(4\partial_-\rho\partial_-\phi-2\partial_-^2\phi)
+{i \over 4}\sum_{j=1}^N
(\psi^{R*}_j\partial_-\psi^R_j-\partial_-\psi^{R*}_j\psi^R_j)
\ ,}
\eqn\ev$$
$$0=T_{+-}=\ephi(2\partial_+\partial_-\phi
-4\partial_+\phi\partial_-\phi-\lambda^2\eprho)
+{2 \over g_A^2}\e^{2\phi-2\rho}F_{+-}^2
\ . \eqn\evi$$
The Maxwell equations are
$$\eqalign{0&=-{8 \over g_A^2}\partial_+(\ephirho F_{+-})
+\half\sum_{j=1}^N\psi^{L*}_j\psi^L_j\ ,\cr
0&={8 \over g_A^2}\partial_-(\ephirho F_{+-})
+\half\sum_{j=1}^N\psi^{R*}_j\psi^R_j\ .}
\eqn\evii$$
We also obtain dilaton and matter equations of motion,
$$0=-4\partial_+\partial_-\phi+4\partial_+\phi\partial_-\phi
+2\partial_+\partial_-\rho+\lambda^2\eprho
-{2 \over g_A^2}\e^{4\phi-2\rho} F_{+-}^2
\ , \eqn\eviii$$
$$0=\partial_-\psi^L_j\ , \ \ \
0=(\partial_+-iA_+)\psi^R_j \ .\eqn\eix$$

By using Equations \evi\ and \eviii , we find that the current $J^\mu$ defined
in Equation \qii\ is conserved.
Therefore we can fix the residual gauge symmetry of
reparametrization by choosing the condition, $\phi-\rho=0$.
In the following, we only consider the solutions with $\psi^R_j=0$.
Then the general solutions of Equations \ev\ - \eix\ are given by,
$$\eqalign{\psi^L_j&=\psi^L_j(x^+)\ ,\cr
A_+&=-{g_A^2 \over 4}x^-\int^{x^+}dy^+\sum_{j=1}^N\psi^{L*}_j(y^+)\psi^L_j(y^+)
\ ,\cr
\ephi&=\erho={M \over \lambda}-\lambda^2x^+x^-
+{g_A^2 \over 8}x^-
\int^{x^+}dy^+(\int^{y^+}dz^+\sum_{j=1}^N\psi^{L*}_j(z^+)\psi^L_j(z^+))^2\cr
&\ \ \ \
-{i \over 4}\int^{x^+}dy^+
\int^{y^+}dz^+\sum_{j=1}^N(\psi^{L*}_j(z^+)\partial_+\psi^L_j(z^+)
-\partial_+\psi^{L*}_j(z^+)\psi^L_j(z^+))\ .}
\eqn\ex$$

Now we consider solutions with $\psi^L_j=0$.
A special solution describing a dilaton vacuum is given by,
$$A_+=0, \ \ \ephi=\erho=-\lambda^2x^+x^-\ .
\eqn\exa$$
We have more general solutions,
$$\eqalign{F_{+-}&=C \ \ ({\rm constant})\ .\cr
\ephi&=\erho={M \over \lambda}-\tilde\lambda^2x^+x^-\ .}
\eqn\exi$$
Here ${M \over \lambda}$ is an integration constant
and $\tilde\lambda$ is defined by
$$\tilde\lambda^2\equiv \lambda^2-{2C^2 \over g_A^2}\ . \eqn\exiii$$
The solutions \exi\ tell that the metric has a singularity when
${M \over \lambda}-\tilde\lambda^2x^+x^-=0$ and there
appear horizons $x^+x^-=0$. Therefore, if $C\neq 0$, the solutions \exi\
describe charged black holes.
%
The structure of space-time is similar
to that of the Schwarzschild black holes and simple in contrast to
the \RN\ black hole solutions in four dimensions.
Note that the singularity vanishes when $\tilde\lambda^2=0$
when we fix $M$ to be finite,
although the solution corresponds to a massive\foot{
Here ``massive" means that ${M \over \lambda}$ does not vanish.}
and charged object.
This solution could be a natural analogue of extremal
\RN\ black hole solution.
In fact this solution has a vanishing temperature.
The temperature can be found by changing the coordinates
$$\eqalign{x^+&={1 \over \tilde\lambda}\sqrt{{M \over \lambda}}
\e^{\tilde\lambda t }\sinh \tilde\lambda r\cr
x^-&=-{1 \over \tilde\lambda}\sqrt{{M \over \lambda}}
\e^{-\tilde\lambda t }\sinh \tilde\lambda r}
\eqn\exii$$
and Wick rotating $t \longrightarrow i\tau$.
The resulting metric is given by
$$ds^2=dr^2+\tanh^2\tilde\lambda r\, d\tau^2\ .\eqn\eixx$$
The geometry approaches to Euclidean flat space-time when $r$ goes to
$\infty$, $ds^2\sim dr^2+d\tau^2$.
On the other hand, the metric has a following form when $r\rightarrow 0$,
$$ds^2\sim dr^2+\tilde\lambda^2 r^2 d\tau^2 \ . \eqn\exx$$
Since the metric is not singular if and only if $\tau$ has a period of
${2\pi \over \tilde\lambda}$, the temperature $T$ is given by,
$$T={\tilde\lambda \over 2\pi}\ .\eqn\exxi$$
This equation \exxi\ tells that the extremal solutions, which correspond
to $\tilde\lambda^2=0$, have a vanishing temperature. This result should
compare with CGHS black holes, which have a common non-vanishing
temperature ${\lambda \over 2\pi}$.
When $\tilde\lambda^2<0$, the temperature is imaginary and
the naked singularity appears.

If we fix $\tilde M=\tilde\lambda^2 M$ to be finite
and redefine the coordinates $x^\pm$ by,
$$x^+\rightarrow x^++{\alpha \tilde M \over \lambda \tilde\lambda^2}\ , \ \
x^-\rightarrow x^-+\beta+{1 \over \alpha\tilde\lambda^2}\ , \eqn\exxa$$
we obtain the following metric,
in the limit of $\tilde\lambda^2\rightarrow 0$,
$$\ephi=\erho={1 \over \alpha}x^++\alpha \lambda \tilde M x^-
+\beta\ .\eqn\exxb$$
The curvature is given by,
$$\eqalign{R&=8\erho\partial_+\partial_-\rho\cr
&={4\lambda \tilde M \over
{1 \over \alpha}x^++\alpha \lambda \tilde M x^-+\beta}\ ,}
\eqn\exxc$$
and we find that there appears a singularity when
${1 \over \alpha}x^++\alpha \lambda \tilde M x^-+\beta=0$.

The solutions \ex\ can also describe charged matter (chiral fermions)
collapsing into black holes. For example, we can consider a charged
shock wave which is given by,
$$\eqalign{{i \over 4}\sum_{j=1}^N(\psi^{L*}_j(x^+)\partial_+\psi^L_j(x^+)
-\partial_+\psi^{L*}_j(x^+)\psi^L_j(x^+))&=a\delta(x^+-x^+_0)\ ,\cr
{g_A^2 \over 4}\sum_{j=1}^N\psi^{L*}_j(x^+)\psi^L_j(x^+)&=b\delta(x^+-x^+_0) \
{}.
}
\eqn\exxii$$
Then the solution is given by,
$$\eqalign{x^+<x^+_0\ , \ \ F_{+-}&=C\ ,\cr
\ephi&=\erho={M \over \lambda}-(\lambda^2-{2C^2 \over g_A^2})x^+x^-\ ,\cr
x^+>x^+_0\ , \ \ F_{+-}&=C+b\ ,\cr
\ephi&=\erho={M \over \lambda}+ax^+_0
+{2Bax^+_0 \over g_A^2\tilde\lambda'^2}\cr
&-\tilde\lambda'^2
(x^++{2Bx^+_0 \over g_A^2\tilde\lambda'^2})
(x^-+{a \over \tilde\lambda'^2})\ .}
\eqn\exxiii$$
Here $\tilde\lambda'$ and $B$ are defined by,
$$\tilde\lambda'^2\equiv\lambda^2-{2(C+b)^2 \over g_A^2}\ , \ \
B\equiv (2Cb+b^2)\ . \eqn\exxiiia$$
When $\tilde\lambda'=0$, the metric \exxiii\ in case $x^+>x^+_0$ correspnds
to the metric in Equation \exxb .
Note that there appears a singularity even in case $\tilde\lambda'=0$.

Equation \exxiii\ tells that the event horizon when $x^+>x^+_0$ is given by
$$(x^++{2Bx^+_0 \over g_A^2\tilde\lambda'^2})
(x^-+{a \over \tilde\lambda'^2})=0\ .
\eqn\exxiv$$
Note that not only future horizon corresponding to
$x^-=-{a \over \tilde\lambda'^2}$, but past horizon, where
$x^+=-{2Bx^+_0 \over g_A^2\tilde\lambda'^2}$, is shifted by the
shock wave. In the original paper by 't Hooft\cref\v the shift of
only future horizon was discussed in four dimensional black holes.
The shift of the past horizon, which is observed in this paper,
would suggest that charged particles collapsing into \RN\ black hole
in four dimensions could shift the past horizon.
The incoming matters cause the change of the charge distribution,
which would shift the past horizon.

\chapter{Semi-classical Analysis of Hawking Evaporation}

In the original paper by CGHS, they have found
that the Hawking radiation and
the back reaction of the metric can be described by adding correction
terms to the classcal equation of motion.
The correction terms come from the conformal anomaly.
Now, since we have massless fermions, the terms which come from
the chiral anomaly should be also added.
Furthermore, by following to \RST\cref\xvii we also add a term $S_R$
corresponding to $\delta S$ in Equation \qi ,  to preserve
the conservation of the current $J^\mu$ in Equation \qii .
Then the total quantum action $S_q$
is given by,
$$\eqalign{S_q&=S_c+S_\rho+S_\chi+S_R\ ,\cr
S_\rho&=\pint\sqg\{-\half(\nabla Z)^2+\sqrt{{N \over 48}}ZR\}\ ,\cr
S_\chi&=\pint\{\half\sqg(\nabla Y)^2
+\sqrt{{N \over 2}}Y\epsilon^{\mu\nu}F_{\mu\nu}\}\ ,\cr
S_R&=-{N \over 96\pi}\int d^2x\sqg \phi R\ .
}\eqn\exxv$$
Here $S_c$ is a classical action in Eq.\ei.
By adding new degrees of freedom $Z$ and $Y$, we write the actions in local
forms.
We have neglected the one loop and higher order contributions from the
reparametrization ghosts, Liouville mode and dilaton field by
assuming that the number of fermions $N$ is very large.

In the following, we only consider the solutions where fermions $\psi^{L,R}_j$
vanish.
By choosing the gauge fixing conditions \eii\ and \eiib ,
the classical equations of motion \ev\ - \evii\ are modified as follows,
$$\eqalign{0=T_{++}=&(\ephi+{N \over 96})
(4\partial_+\rho\partial_+\phi-2\partial_+^2\phi)
+\half\partial_+Z\partial_+Z-\sqrt{{N \over 12}}\partial_+\rho\partial_+Z\cr
& \ \ \ +{1 \over 2}\sqrt{{N \over 12}}\partial_+^2 Z
-\half\partial_+Y\partial_+Y\ , \cr
0=T_{--}=&(\ephi+{N \over 96})
(4\partial_-\rho\partial_-\phi-2\partial_-^2\phi)
+\half\partial_-Z\partial_-Z-\sqrt{{N \over 12}}\partial_-\rho\partial_-Z\cr
& \ \ \ +{1 \over 2}\sqrt{{N \over 12}}\partial_-^2 Z
-\half\partial_-Y\partial_-Y\ , }
\eqn\exxvi$$
$$\eqalign{0=&T_{+-}=\ephi(2\partial_+\partial_-\phi
-4\partial_+\phi\partial_-\phi-\lambda^2\eprho)
+{2 \over g_A^2}\e^{2\phi-2\rho}F_{+-}^2 \cr
&-\sqrt{{N \over 48}}\partial_+\partial_-Z
+{N \over 48}\partial_+\partial_-\phi
\ ,} \eqn\exxvii$$
$$\eqalign{0&=-{8 \over g_A^2}\partial_+(\ephirho F_{+-})
-\sqrt{2N}\partial_+Y \ ,\cr
0&=-{8 \over g_A^2}\partial_-(\ephirho F_{+-})
-\sqrt{2N}\partial_-Y
\ . }\eqn\exxviii$$
The dilaton equation of motion \eviii\ is modified to be,
$$\eqalign{0=&-4\partial_+\partial_-\phi+4\partial_+\phi\partial_-\phi
+2\partial_+\partial_-\rho+\lambda^2\eprho \cr
&-{2 \over g_A^2}\e^{4\phi-2\rho} F_{+-}^2
+\epphi{N \over 48}\partial_+\partial_-\rho
\ ,}\eqn\exxviiia$$
We also have $Z$ and $Y$ equations,
$$0=-2\partial_+\partial_-Z
+\sqrt{{N \over 3}}\partial_+\partial_-\rho \ ,\eqn\exxixa$$
$$0=2\partial_+\partial_-Y+\sqrt{2N}F_{+-} \ .\eqn\exxixb$$

By using Equations \exxvii\ and \exxviiia , we find
$\partial_\mu\partial^\mu(\phi-\rho)=0$ and the current $J^\mu$
defined in \qii\ is conserved.
Therefore we can impose the residual gauge condition $\rho=\phi$.
Under this gauge condition, the modified Maxwell equation \exxviii\ can be
solved with respect to $Y$ as follows
$$\sqrt{2N}Y=-{8 \over g_A^2}F_{+-}+c\ . \eqn\exxx$$
Here $c$ is a constant.
By substituting this equation \exxx\ into Equation \exxixb ,
we obtain,
$$0=\partial_+\partial_-F_{+-}-{g_A^2 N \over 8}F_{+-} \ .
\eqn\exxxi$$
This equation tells that electromagnetic fields become massive due to
the screening effect by fermion loops.
If we consider a situation where an incoming shock wave \exxii\ carries
charge, the boundary are given by,
$$\eqalign{
F_{+-}&=C \ \ {\rm when}\ \ x^+=x^+_0\ , \cr
\partial_+F_{+-}&=0 \ \ {\rm when}\ \ x^+=x^+_0\ {\rm and}\ x^-=\txz\ .}
\eqn\qqi$$
The latter condition is imposed to fix the constant of integration.\foot{
Since the velocity of massless matter is larger than that of electro-magnetic
fields, which is massive, the trajectory of the matter makes a shock wave
of electro-magnetic field like Cherenkov radiation. The parameter $\txz$ is
related to the strength of the shock wave.}
The solution of Equation \exxxi\ with the boundary condition \qqi\
is given by,
$$\eqalign{
F_{+-}=&CJ_0(p) \ \ \ \ \
(x^-<\txz)\cr
=&CI_0(q) \ \ \ \ (x^->\txz)                                            \ .}
\eqn\qqii$$
Here $J_n$ is a Bessel function and $I_n$ is a modified Bessel function:
$$\eqalign{J_n(z)&=({z \over 2})^n\sum_{m=0}^\infty
{(-1)^m(z/2)^{2m} \over m!\Gamma (m+n+1)}\ , \cr
I_n(z)&=({z \over 2})^n\sum_{m=0}^\infty
{(z/2)^{2m} \over m!\Gamma (m+n+1)}\ ,}\eqn\qqiii$$
and $p$ and $q$ are defined by,
$$p\equiv \sqrt{g_A^2N(\txz-x^-)(x^+-x^+_0) \over 2}\ , \ \
q\equiv \sqrt{-{g_A^2N(\txz-x^-)(x^+-x^+_0) \over 2}}\ .
\eqn\qqx$$
This screening effect makes the field strength $F_{+-}$ decrease
when $x^-<\txz$ \ie , in the asymptotic region and
increase near when $x^->\txz$.
As we will see later, the singularity can disappear in a finite proper
time due to this screening effect.

Equation \exxixa\ can be also solved with respect to $Z$ as follows,
$$Z=\sqrt{{N \over 12}}\rho+r^+(x^+)+r^-(x^-) \ .\eqn\exxxii$$
By using this equation and the gauge condition $\rho=\phi$,
Equation \exxvi\ can be written as follows,
$$\eqalign{0&=\partial_+^2(\ephi+{N \over 48}\phi)+t^+(x^+)
-{1 \over 4N}\{\partial_+(-{8 \over g_A^2}F_{+-})\}^2\cr
0&=\partial_-^2(\ephi+{N \over 48}\phi)+t^-(x^-)
-{1 \over 4N}\{\partial_-(-{8 \over g_A^2}F_{+-})\}^2
\ . }
\eqn\exxxiii$$
Here $t^\pm(x^\pm)$ is defined by
$$t^\pm(x^\pm)\equiv \half(\partial_\pm r^\pm)^2
+{1 \over 4}\sqrt{{N \over 3}}\partial_\pm^2r^\pm \ . \eqn\exxxiv$$
Since $r^\pm$ is a solution of the equation motion \exxixa ,
$t^\pm(x^\pm)$ represents dynamical degrees of freedom.
Furthermore we obtain the following equations by using Equations
\exxvii , \exxviiia, and \exxxii\ and the gauge condition $\phi-\rho=0$,
$$0=\partial_+\partial_-(\ephi+{N \over 48}\phi)
+\lambda^2-{2 \over g_A^2}F_{+-}^2\ .
\eqn\exxxivb$$
The classical solution describing the dilaton vacuum \exa\ satisfies Equations
\exxxi\ and \exxxivb .
If we require that the solution also satisfies \exxxiii ,
we find that $t^\pm(x^\pm)$ in Equation \exxxiii\ should be given by,
$$t^\pm(x^\pm)=-{N \over 96 (x^\pm)^2} \ .\eqn\exxxv$$
Although $t^\pm(x^\pm)$ can be an arbitrary function of $x^\pm$ in general,
we assume Equation \exxxv\ for a while.
The contribution of the energy-momentum tensor of massless matter
fields can effectively shift $t^\pm(x^\pm)$.

Equations \exxxi , \exxxiii\ and \exxxivb\ can be easily solved.
A static solution is given by,\foot{
There is an ambiguity for the definition of $\ln (-\lambda^2x^+x^-)$
when $\lambda^2x^+x^->0$.}
$$\eqalign{
F_{+-}=&CJ_0(p_0) \ \ (x^-<0)\cr
=&CI_0(q_0) \ \ (x^->0)\ ,\cr
\om=&{M \over \lambda}-\lambda^2x^+x^--{N \over 96}\ln (-\lambda^2x^+x^-)
+{1 \over N}({2C \over g_A^2})^2L_0(x^+,x^-)
}\eqn\qqiv$$
Here $L_0(x^+,x^-)$ is defined by
$$\eqalign{L_0(x^+,x^-)=&-p_0^2(J_0(p_0)^2+2J_1(p_0)^2-J_0(p_0)J_2(p_0))\ ,
\ \ \ (x^-<0)\cr
=&q_0^2(I_0(q_0)^2-2I_1(q_0)^2+I_0(q_0)I_2(q_0))\ , \ \ \
(x^->0)\ }
\eqn\qqv$$
and $p_0$ and $q_0$ are defined by,
$$p_0\equiv \sqrt{-{g_A^2Nx^-x^+ \over 2}}, \ \
q_0\equiv \sqrt{g_A^2Nx^-x^+ \over 2} \ .
\eqn\qqvi$$
The solution \qqiv\ does not depend on ``time'' $t$ when we change the
coordinate by,
$$x^+=\e^{g(t)}f(r)\ , \ \ \ x^-=\e^{-g(t)}f(r)\ .\eqn\qqvii$$
Here $g$ and $f$ are arbitrary functions.

We can also obtain the solutions describing the matter shock wave
collapsing into the dilaton vacuum.
When we consider the shock wave whose trajectory is $x^+=x^+_0$,
$F_{+-}$ is given by Equation \qqii\ and $\om$ is given by
$$\om={M \over \lambda}-\lambda^2x^+(x^-+{M \over \lambda^3x^+_0})
-{N \over 96}\ln (-\lambda^2x^+x^-)
+{1 \over N}({2C \over g_A^2})^2L(x^+,x^-)\ .\eqn\qqviii$$
Here $L(x^+,x^-)$ is defined by
$$\eqalign{L(x^+,x^-)=&-p^2(J_0(p)^2+2J_1(p)^2-J_0(p)J_2(p))\ , \
(x^-<\txz)\cr
=&q^2(I_0(q)^2-2I_1(q)^2+I_0(q)I_2(q))\ . \
(x^->\txz)}
\eqn\qqvix$$

\chapter{The Fate of Singularity and Apparent Horizon}

In this section, we discuss the behavior of curvature singularity and
apparent horizon by using the exact solution \qqviii .
In this section we assume $\tilde\lambda^2\equiv \lambda^2-{2C^2 \over g_A^2}$
is positive.
The main results are following:

\item{1.} When $\lambda^3x_0^+\txz>-M$, the essential behavior
is similar to that of Callan, Giddings, Harvey and
Strominger's dilatonic black hole theory modified in the paper
by \RST\pref\xvii
The singularity formed by the collapsing matter emerged naked.
\item{2.} When $\lambda^3x_0^+\txz<-M$,
the singularity and the apparent horizon disappear in a finite proper time.
Furthermore, there appears a discontinuity in the metric on the
trajectory
of the collapsing matter
($x=x_0^+$, $x^->-{1 \over \lambda^2 x^+_0}{N \over 96}$).

The curvature $R=8\erho\partial_+\partial_-\rho$ become singular when
$\ephi={N \over 96}$ since
$$R\propto (\ephi-{N \over 96})^{-1}\ .\eqn\qqxi$$
By using Equation \qqviii , we find the line of the singularity is given by,
$$\lambda^2x^+x^-+{M \over \lambda x^+_0}(x^+-x^+_0)
+{N \over 96}\ln (-\lambda^2x^+x^-)
+{N \over 96}(1-\ln {N \over 96})
={1 \over N}({2C \over g_A^2})^2L(x^+,x^-)\ .\eqn\qqxii$$
A point
on the trajectory
$(x^+,x^-)=(x_0^+, -{1 \over \lambda^2 x^+_0}{N \over 96})$
satisfies Equation \qqxii.

We now discuss about the solution of the Equation \qqxii\ when
$x^+$ is large enough. It is convenient to divide the both side of
Equation \qqxii\ by $x^+$ and define functions ${\cal R}(x^-)$ and
${\cal L}(x^-)$ by
$$\eqalign{
{\cal R}(x^-)&={1 \over x^+}\{\lambda^2x^+x^-
+{M \over \lambda x^+_0}(x^+-x^+_0)+{N \over 96}\ln (-\lambda^2x^+x^-)
+{N \over 96}(1-\ln {N \over 96})\} \ ,\cr
{\cal L}(x^-)&={1 \over Nx^+}({2C \over g_A^2})^2L(x^+,x^-)\ .}\eqn\qqxiii$$
Then the singularity line is given by
${\cal R}(x^-)={\cal L}(x^-)$.
When $x^+\rightarrow +\infty$, the curve
${\cal C}^{\cal R}=\{(x^-, y)\, |\, y={\cal R}(x^-)\}$
approaches to the followng curve
$$\eqalign{{\cal C}^{\cal R}\LA
{\cal C}^{\cal R}_\infty&=\{(x^-, y)\,|\, y
=\lambda^2(x^-+{M \over \lambda^3x^+_0})\ ,\
\ x^-<0\}\cr
&\oplus\{(x^-, y)\,|\,y<{M \over \lambda x^+_0}\ ,
\ \ x^-=0\}}\eqn\qqxiv$$
On the other hand, the curve
${\cal C}^{\cal L}=\{(x^-, y)|y={\cal L}(x^-)$ approaches to the following
curve
$${\cal C}^{\cal L}\LA
{\cal C}^{\cal L}_\infty=\{(x^-, y)\,|\,y =0\ ,\ \ x^-<\txz\}
\oplus\{(x^-, y)\,|\, y\geq 0\ ,\ \ x^-=0\}\eqn\qqxv$$
When $\lambda^3x_0^+\txz>-M$, there appear two solutions satisfying
the equation ${\cal R}(x^-)={\cal L}(x^-)$
in the limit $x^+\rightarrow\infty$,
$$\eqalign{
x^-&=-{M \over \lambda^3x^+_0}\ ,\cr
x^-&=0\ (\txz>0)\ \ {\rm or}\ \ x^-=\txz\ \ (\txz<0)\ .}
\eqn\qqxvi$$
Note that there appears two lines of singularities.

When $\lambda^3x_0^+\txz<-M$, there is not any solution
satisfying the equation ${\cal R}(x^-)={\cal L}(x^-)$ when $x^+=\infty$.
This tells that the curvature singularities disappear in a finite proper time.
The disappearance occurs in a weak coupling region $\ephi<{N \over 96}$ and
the semi-classical analysis given here is reliable.
The collapsing matter \lq pair creates' two curvature singularities and
these singularities \lq pair annihilate' in a finite proper time
when $\lambda^3x_0^+\txz<-M$, \ie, the line of the singularity makes a loop
in $x^+$-$x^-$ plane.
Outside of the loop, we find $\ephi<{N \over 96}$ in a region $x^+>x^+_0$.
On the other side, in a region $x^+<x^+_0$ where the solution describes
the dilaton vacuum \exa , $\ephi>{N \over 96}$ just below the trajectory
of collapsing matter $x^+=x^+_0-0$ when
$x^->-{1 \over \lambda^2 x^+_0}{N \over 96}$.
Therefore there appears a discontinuity on the trajectory
$x^+=x^+_0$, $x^->-{1 \over \lambda^2 x^+_0}{N \over 96}$
although $\om$ is continuous.
The discontinuity would be a signal of topology change.

Although the curvature singularity appears at
$(x^+,x^-)=(x^+_0,-{1 \over \lambda^2 x^+_0}{N \over 96})$,
the singularity disappears on the line $x^+=x^+_0+\epsilon$ ($\epsilon$ is an
infinitesimal constant) if
$$\partial_+{\cal R}(x^-)|_{x^+=x^+_0}<\partial_+{\cal L}(x^-)|_{x^+=x^+_0}
\ , \eqn\qqqi$$
that is,
$$\txz<-{Mg_A^2 \over 2C^2\lambda x^+_0}-{N \over 96\lambda^2x^+_0}\ .
\eqn\qqqii$$
In this case, the curvature singularities appear after a finite proper
time or any singularity does not appear.
In any case, the discontinuity of the metric appears on the trajectory of
the collapsing matter.

Since $\ephi<{N \over 96}$ outside the singularity loop
in a region $x^+>x^+_0$, $\ephi\sim -{N \over 96}\ln (-\lambda^2 x^+x^-)$
when $-\lambda^2 x^+x^- \sim 0$. Then the curvature
$R=8\erho\partial_+\partial_-\rho$ diverges as
$R\sim \{-\lambda^2 x^+x^-\ln (-\lambda^2 x^+x^-)\}^{-1}$.
Therefore there appears
a new line of the singularity when $x^-=0$. The singularity occurs since we
have chosen $t^\pm(x^\pm)$ by Equation \exxxv .
There is an ambiguity to how to choose  $t^\pm(x^\pm)$ since
$t^\pm(x^\pm)$ or $r^\pm(x^\pm)$ in Equation \exxxiv\ can be regarded as a
part of dynamical modes of the field $Z$ in Equation \exxxii .
If we choose $t^-(x^-)$, for example, by
$t^-(x^-)=-{N \over 96 }\half \partial_-^2\ln \{(x^-)^2+\alpha^2\}$
($\alpha$ is a constant), there is no singularity except the singularities
coming from $\ephi={N \over 96}$.
Therefore the singularity which appeared when $-\lambda^2 x^+x^- \sim 0$
would not have a physical meanings.
If we do not change the asymptotic behavior of $t^\pm(x^\pm)$,
the qualitative behavior of the $\ephi={N \over 96}$ singularity line
do not depend on the choice of $t^\pm(x^\pm)$.

We now discuss about the apparent horizon which is defined by,
$\partial_+\phi=0$\pref\vii
By using Equation \qqviii , we find that the apparent horizon is given by,
$$\lambda^2(x^-+{M \over \lambda^3x^+_0})
+{N \over 96x^+}
={1 \over N}({2C \over g_A^2})^2\partial_+L(x^+,x^-)\ .\eqn\qqxvii$$
When $x^+=x^+_0+0$, the solutions which satisfies Equation \qqxvii\
is given by, $x^-=-{1 \over \tilde \lambda^2}({M \over \lambda x^+_0}
+{N \over 96x^+_0}+{2C^2\txz \over g_A^2})$ and $x^-=-\infty$. Note that
there appears two lines of apparent horizons in general.
We define ${\cal R}^A(x^-)$ and
${\cal L}^A(x^-)$ by
$$\eqalign{
{\cal R}^A(x^-)&=\lambda^2(x^-+{M \over \lambda^3x^+_0})
+{N \over 96x^+} \ ,\cr
{\cal L}^A(x^-)&={1 \over N}({2C \over g_A^2})^2\partial_+L(x^+,x^-)\ .}
\eqn\qqxviii$$
The apparent horizon is given by
${\cal R}^A(x^-)={\cal L}^A(x^-)$.
When $x^+\rightarrow +\infty$, the curves
${\cal C}^{A {\cal R}}=\{(x^-, y)\,|\, y={\cal R}^A(x^-)\}$
and ${\cal C}^{A{\cal L}}=\{(x^-, y)\,|\, y={\cal L}^A(x^-)\}$
approach to the following
curves
$$\eqalign{
{\cal C}^{A{\cal R}}\LA & {\cal C}^{A{\cal R}}_\infty=\{(x^-, y)\,|\,
y =\lambda^2(x^-+{M \over \lambda^3x^+_0})\}\cr
{\cal C}^{A{\cal L}}\LA &
{\cal C}^{A{\cal L}}_\infty=\{(x^-, y)\,|\,y =0\ ,\ \ x^-<\txz\}
\oplus\{(x^-, y)\,|\, y\geq 0\ ,\ \ x^-=0\}}\eqn\qqxix$$
When $\lambda^3x_0^+\txz>-M$, there appear two solutions satisfying
the equation ${\cal R}^A(x^-)={\cal L}^A(x^-)$
in the limit of $x^+=\infty$,
$$x^-=-{M \lambda^3x^+_0}\ ,\ \ \ x^-=\txz\ .\eqn\qqxx$$
In this case, the behaviors of the lines of singularities and apparent
horizons are qualitatively similar to that of Callan, Giddings, Harvey and
Strominger's dilatonic black hole theory modified in the paper
by \RST\refmark\xvii and the singularity emerges naked.
When $\lambda^3x_0^+\txz<-M$, there is not any solution
satisfying the equation ${\cal R}^A(x^-)={\cal L}^A(x^-)$
in the limit of $x^+=\infty$
and the apparent horizons disappear in a finite proper time.

\chapter{Summary and Discussion}

We have analyzed exactly solvable semi-classical theory of
two dimensional dilatonic gravity with electromagnetic interactions.
We added the term which changes the kinetic term and keeps the current
$J^\mu=\partial^\mu (\phi-\rho)$ to conserve.
The theory contains massless fermions as matter fields and
there appear the quantum corrections including chiral anomaly.
The screening effect due to the chiral anomaly has a tendency to
cloak the singularity.
When $\lambda^3x_0^+\txz<-M$,
the singularity coming from $\ephi={N \over 96}$
and the apparent horizon disappear in a finite proper time.
Furthermore, there appears a discontinuity in the metric on the trajectory
of the collapsing matter, which would be a signal of topology
change.

One of the problems is whether the disappearence of the singularities
observed here is independent of the details of the models since
many semi-classical models have been proposed\pref{\xviii - \xvii}
We have also analyzed the model without $\phi R$ term in \qi , which
corresponding to the original model by CGHS.
By assuming that
$\tilde\lambda^2\equiv \lambda^2-{2C^2 \over g_A^2}$ is small compared
to $\lambda^2$, we have investigated the line $x^-=f(x^+$ where $\phi$ is a
constant (on the singularity $\phi=-{1 \over 2}\ln {N \over 96}$) and
we find
$$f(x^+)=f_0-f_1{\tilde\lambda^2 \over \lambda^2}(x^+-x^+_0)
+f_2(x^+-x^+_0)^2+{\cal O}((x^+-x^+_0)^3)\ .\eqn\qqqq$$
Here $f_i$'s are order one quantities with respect to
${\tilde\lambda^2 \over \lambda^2}$ and positive if
$\txz \ll -{N \over 24\lambda^2x^+_0}$.
Therefore there exists a point $x^+=\tilde x^+_0$ which satisfies
$\tilde x^+_0-x^+_0={\cal O}({\tilde\lambda^2 \over \lambda^2})$,
$\tilde x^+_0-x^+_0>0$ and $f'(x^+)=0$.
This tells that there is a turning point and we have found that
the screening effect due to the chiral anomaly has a tendency to
cloak the singularity.

\ack{
We acknowledge K. Odaka and A. Sugamoto for discussions.
We are also indebted to S. Odake and T. Tada for useful informations.
The research of I.O. is supported by the Japan Society for the Promotion
of Science.}

\refout

\bye